\documentclass[%
 reprint,
superscriptaddress,
 amsmath,amssymb,
 aps,
]{revtex4-1}

\usepackage{graphicx}
\usepackage{dcolumn}
\usepackage{bm}
\usepackage{amsmath}
\usepackage{listings}
\usepackage{physics}
\usepackage[dvipsnames]{xcolor}
\usepackage{graphicx}
\usepackage{dcolumn}
\usepackage{bm}
\usepackage{units}
\usepackage[utf8]{inputenc}
\usepackage[T1]{fontenc}
\usepackage[colorinlistoftodos]{todonotes}

\begin{document}

\preprint{APS/123-QED}

\title{Strictly Non-Adiabatic Quantum Control of the Acetylene Dication Using an Infrared Field}

\author{Chelsea Liekhus-Schmaltz}
\email{cliekhus@stanford.edu}
\affiliation{ Stanford PULSE Institute, SLAC National Accelerator Laboratory\\
2575 Sand Hill Road, Menlo Park, CA 94025}
\affiliation{Department of Physics, Stanford University, Stanford, CA 94305}

\author{Xiaolei Zhu}
\affiliation{ Stanford PULSE Institute, SLAC National Accelerator Laboratory\\
2575 Sand Hill Road, Menlo Park, CA 94025}

\author{Gregory A. McCracken}
\affiliation{ Stanford PULSE Institute, SLAC National Accelerator Laboratory\\
2575 Sand Hill Road, Menlo Park, CA 94025}
\affiliation{Department of Applied Physics, Stanford University, Stanford, CA 94305}

\author{James P. Cryan}
\affiliation{ Stanford PULSE Institute, SLAC National Accelerator Laboratory\\
2575 Sand Hill Road, Menlo Park, CA 94025}

\author{Todd Martinez}
\affiliation{ Stanford PULSE Institute, SLAC National Accelerator Laboratory\\
2575 Sand Hill Road, Menlo Park, CA 94025}
\affiliation{Department of Chemistry, Stanford University, Stanford, CA 94305}

\author{Philip H. Bucksbaum}
\affiliation{ Stanford PULSE Institute, SLAC National Accelerator Laboratory\\
2575 Sand Hill Road, Menlo Park, CA 94025}
\affiliation{Department of Physics, Stanford University, Stanford, CA 94305}
\affiliation{Department of Applied Physics, Stanford University, Stanford, CA 94305}

\date{\today}

\begin{abstract}
We demonstrate the existence of a strictly non-adiabatic control pathway in deprotonation of the acetylene dication.  This pathway is identified experimentally by measuring a kinetic energy shift in an ion coincidence experiment.  We use a TDSE simulation to identify which properties most strongly affect our control.  We find that resonant control around conical intersections is limited by the speed of non-adiabatic dynamics.  
\end{abstract}

\maketitle

\section{\label{sec:Introduction}Introduction}

The Born-Oppenheimer Approximation (BOA) framework forms the backbone of our current understanding of quantum dynamics in molecules \cite{klessinger_excited_1995,born_zur_1927}.  
In the BOA framework the molecular wavefunction is separated into nuclear and electronic parts.  
Internuclear motion then proceeds on an electronic potential energy surface (PES) in nuclear coordinate space.  
This is valid when the surfaces representing different electronic states are well separated; but when potential surfaces approach each other, the BOA framework must be modified and motion is no longer described by a single PES.  
The resulting dynamics are referred to as non-adiabatic and are characterized by motion that departs from the adiabatic pathways.  

Non-adiabatic motion is often described in the adiabatic electronic basis set, where the electronic states are calculated by diagonalization at fixed nuclear positions. 
Points of degeneracy between two adiabatic surfaces in the BOA framework are called `Conical Intersections' (CIs), and they are coupled by a non-adiabatic coupling term that grows to infinity at the point of degeneracy \cite{worth_beyond_2004,yarkony_diabolical_1996,levine_isomerization_2007}.  
CIs have been important for predicting molecular motion in many systems.  
Some striking examples include isomerization in ethylene and ring opening in 1,3-Cyclohexadiene \cite{levine_isomerization_2007,kim_control_2012}, which are important prototypes for important photochemicals such as Vitamin D \cite{barton_selective_2008,pitts_photochemistry_2007}. 
The non-adiabatic coupling term has the following form:
\begin{equation}
    \Lambda_{i,j} \sim \frac{\bra{\Phi_j}(\overrightarrow{\nabla}_N{\hat{H}}_{el})\ket{\Phi_i}}{V_i-V_j}\cdot\overrightarrow{\nabla}_N,
\end{equation}
where $\Lambda_{i,j}$ is the non-adiabatic coupling, $\ket{\Phi_i}$ are the electronic eigenstates, $\overrightarrow{\nabla}_N$ indicates derivatives with respect to nuclear degrees of freedom, $\hat{H}_{el}$ is the electronic Hamiltonian for fixed nuclei, and $V_j$ are the potentials for each adiabatic electronic state.
Critically, we see that the non-adiabatic coupling scales with the speed of the wavepacket as well as the energy splitting between states. 

Several efforts have been made to control non-adiabatic motion by applying an external light field.  
Some schemes couple population between different electronic states to introduce new pathways, while others use coherent methods that interact with the phase of the wavefunction \cite{zhou_probing_2012,kim_control_2012,abe_optimal_2005,lim_experimental_2010,arnold_control_2018,liekhus-schmaltz_coherent_2016}.  
Still others make use of Stark shifted electronic states to modify the decay pathways of non-adiabatic motion \cite{wolter_ultrafast_2016,townsend_Stark_2011}.
In all of these methods the Hamiltonian must include a transition dipole coupling term for the light interaction in addition to the non-adiabatic coupling.  
Therefore, critical to the success of these methods, and designing new ones, is an understanding of how these two coupling terms interact \cite{kim_ab_2015}.  
In Ref.~\cite{liekhus-schmaltz_coherent_2016}, we proposed a control method that would involve a direct interaction between non-adiabatic coupling and dipole coupling, called kinetic energy control.  
In this paper, we observe the kinetic energy control pathway (KECP) in deprotonation in the acetylene dication by measuring the effect a control field has on the kinetic energy release (KER).  

\begin{figure}
\includegraphics[width=\columnwidth]{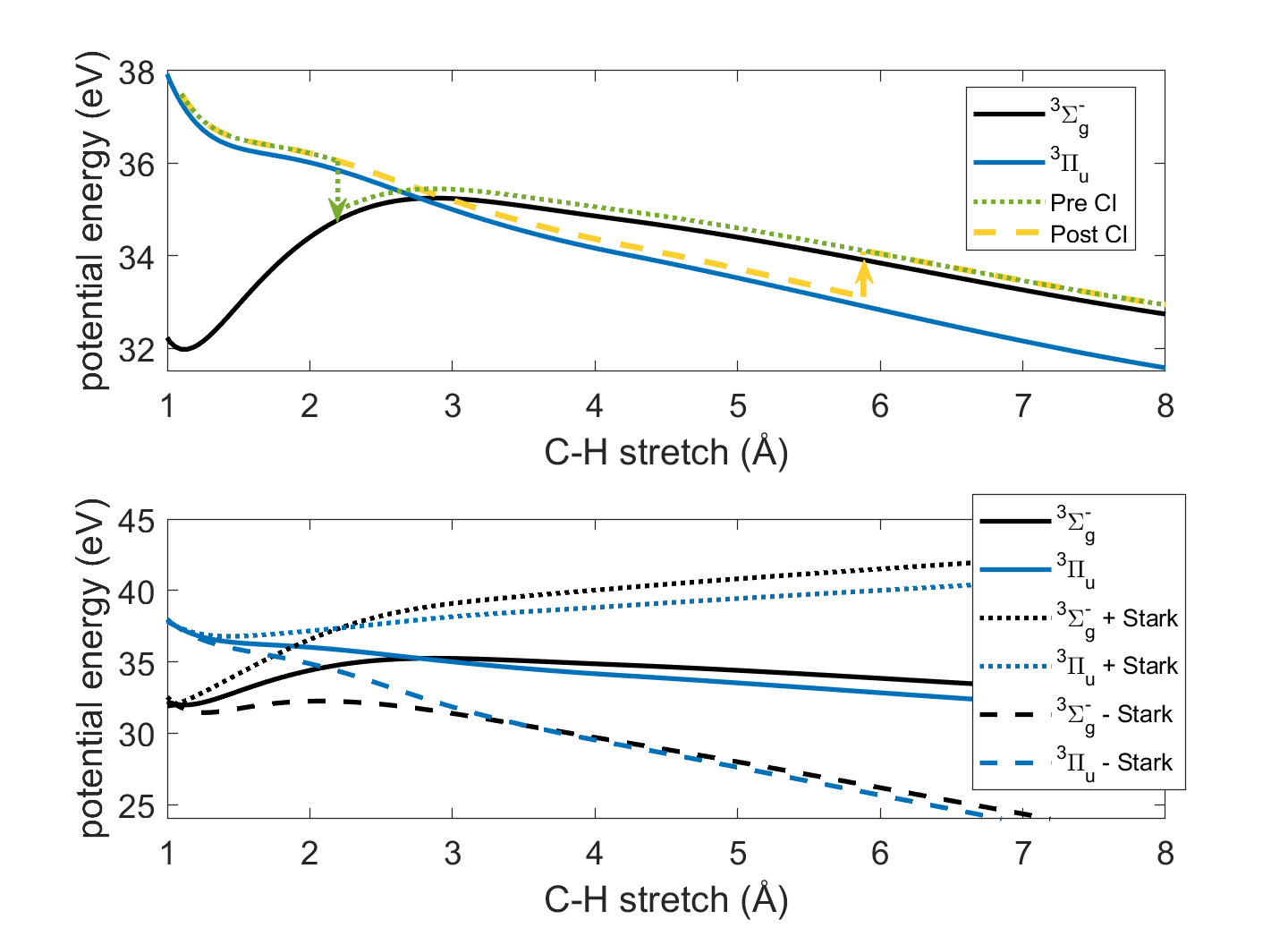}
\caption{The potential energy of the acetylene dication along the C-H stretch calculated using Fractional-Occupation Molecular Orbital Complete-Active-Space Connguration Interaction
(FOMO-CASCI) level of theory.  
The upper panel shows the field free $^3\Sigma_g^+$ state as a solid black curve.  
The field free $^3\Pi_u$ is the blue curve.  
The green dotted line is the kinetic energy control pathway.  
This pathway will result in a KER of 4 eV.  
Dipole coupling can also occur after the CI, as shown as the yellow dashed line.  
This pathway results in a KER of 5 eV.  
The lower panel shows the dynamic Stark shifted potential curves.  
The diagonal terms of the dipole coupling matrix represent a Stark shift, and can therefore affect the position of the CI.\cite{wolter_ultrafast_2016}
This can change the population distribution after interacting with a field.}
\label{fig:potentials}
\end{figure}

Deprotonation in the acetylene dication occurs in 15 fs via the $^1\Pi_u$ and the $^3\Pi_u$ states.
A laser can couple these states to the $^1\Sigma_g^+$ and $^1\Delta_g$ states and the $^3\Sigma_g^-$ state respectively \cite{alagia_angular_2012,jiang_ultrafast_2010,jacobson_acetylene_2000,flammini_vinylidene_2008,adachi_photoelectronphotoionphotoion_2007, osipov_photoelectron-photoion_2003, osipov_fragmentation_2008, gaire_photo-double-ionization_2014, zyubina_theoretical_2005}.  
The potential energy curves for the $^3\Pi_u$ and $^3\Sigma_g^-$ are shown in the upper portion of Fig.~\ref{fig:potentials}.
The KECP occurs when the wavepacket absorbs a photon before the CI and then non-adiabatically crosses the CI, as shown by the green dotted line in Fig.~\ref{fig:potentials}.  
Other pathways are also possible, such as a field free non-adiabatic decay if the wavepacket follows the $^3\Pi_u$ state (blue solid potential curve), or adiabatic decay if the wavepacket changes from the $^3\Pi_u$ state to the $^3\Sigma_g^-$ state (blue to black solid curves).  The wavepacket can also absorb a photon after the CI, as in the dashed yellow curve.  
Each of these pathways results in different KER, allowing us to measure how the control field affects the decay and therefore interacts with non-adiabatic motion.

\section{Control Experiment}
\label{sec:Experiment}
The goal for our quantum control experiment is to observe a new deprotonation channel that directly involves both non-adiabatic dynamics as well as dipole coupling.  
To do so, we monitor the KER for deprotonation in the acetylene dication with and without an external control field.

The experiment we perform is a two pulse, pump-control experiment.  
We prepare the acetylene dication by doubly ionizing acetylene with a linearly polarized, 200 fs long (FWHM), and approximately $5\times 10^{12}$ W/cm$^2$ 266 nm pulse through multiphoton ionization.  
Control is applied via an overlapping 1300 nm optical pulse that is approximately 200 fs in duration (FWHM), focused to $5\times 10^{12}$ W/cm$^2$ and polarized parallel and perpendicular to the ionizing field.
Since the dynamics take place on a $\approx$ 10 fs timescale and our pulses are $\approx$ 200 fs long, we forego any time resolved measurements and simply overlap the two pulses to apply control.  
We will then measure the KER with a velocity map spectrometer to observe the effect of the control field.  
The KER will show that we were able to induce the KECP, and probability of this pathway will give us information about how these two coupling terms interact.

The control field intensity is set to the maximum intensity such that it cannot ionize the acetylene on its own.
To disentangle the nonlinear increase in ionization when both pump and control pulses are present, we can compare the KER for both perpendicular and parallel control polarization vs.~the ionizing pulse.

The experiment is performed in a coincidence ion velocity-map imaging spectrometer backfilled with acetylene to a total $7*10^{-10}$ torr pressure.  
The momenta of ionized molecular fragments are measured with a Roentdek hex-anode delay-line area detector that records their time of arrival and 2D position in the detector plane.\cite{roentdek_hexanode_2019}
This three-dimensional information is sufficient to completely reconstruct the 3D momentum of each molecular ion fragment as it leaves the interaction region of the spectrometer.

To ensure that we are in the coincidence regime we work at low intensities and pressures so that only 0.3 ions are created per laser pulse.  
We then isolate shots that result in one proton and one C$_2$H$^+$ ion to ensure we are looking at the deprotonation pathway.  
To eliminate most false coincidences, we require that the total momentum sums to near zero.
A similar technique is shown in Ref.~\cite{liekhus-schmaltz_ultrafast_2015}.
This momentum requirement allows us to include coincidences where there is one additional ion besides the two fragments from deprotonation.  

\begin{figure}
\includegraphics[width=\columnwidth]{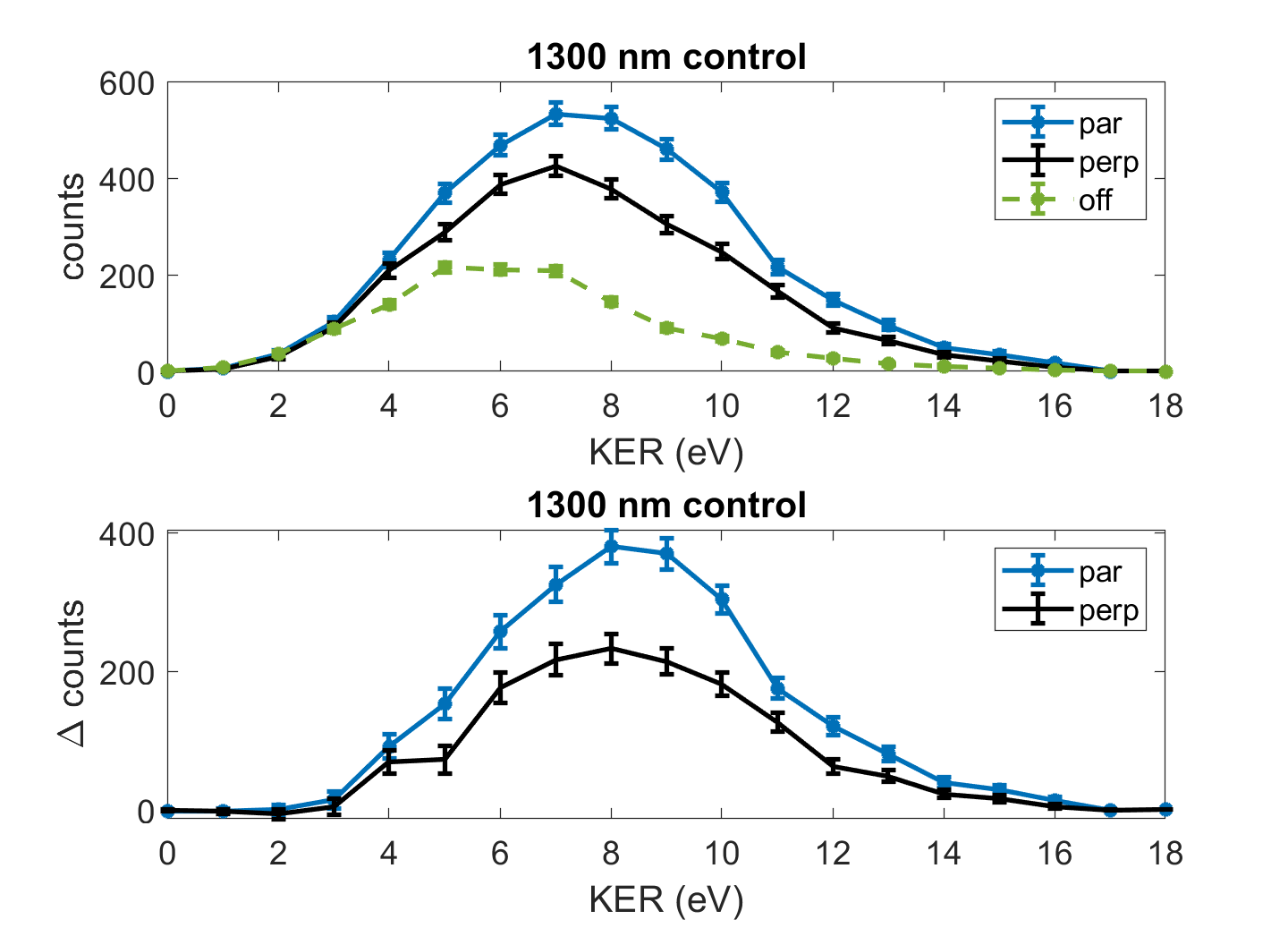}
\caption{The kinetic energy release spectrum for deprotonation.  
The upper panel shows the raw kinetic energy release for the control field off and parallel and perpendicularly polarized conditions relative to the pump field.  
(Labeled as ``off'', ``par'', and ``perp'' respectively.)  
The difference between the control on and control off conditions is shown in the lower panel for both relative polarization.
A shoulder appears in the perpendicular condition at 4 eV that is indicative of the kinetic energy control pathway. }
\label{fig:DP_both}
\end{figure}

We determine the KER of the fragment pair, which indicates the decay pathway as well as energy shifts due to absorption or emission of quanta from the applied field or Stark shifts of the energy levels.  
The KER spectrum is shown in Fig.~\ref{fig:DP_both} for the control field on and off and for parallel and cross polarization settings.  
Figure \ref{fig:DP_both} also shows the change in counts from the control-off case for both control polarizations.  

The most striking observation in Fig.~\ref{fig:DP_both} is the increase in the total number of counts when the control pulse is present.  
This is due to the non-linear increase in ionization from overlapping the two fields. 
To distinguish this effect from our desired quantum control we compare different control polarization.

Since we ionize via multiphoton ionization with 8 photons, the molecular axis of the acetylene molecule is preferentially aligned with the ionizing field.\cite{ibrahim_tabletop_2014} 
Dipole selection rules therefore dictate that dipole coupling between the two states will be strongest with a cross polarized control field.  
In contrast, adding in a control field that is parallel to the ionizing field increases the field strength along the molecular axis more so than a perpendicularly polarized control field, and hence will increase the number of ionized species.  
Processes that are solely due to multiphoton processes will dominate the parallel configuration.  

Polarization dependence of the kinetic energy released by deprotonation shows evidence for new pathways introduced by the 1300 nm control field.
The feature that is due to the KECP is a shoulder at 4 eV KER for perpendicular polarization.  
In contrast to the control off and parallel control conditions, where a KER of 4 eV appears to be due to the broad KER distribution, for perpendicular polarization, there are additional counts resulting in the shoulder.
To confirm this assignment we use the PESs calculated in Fig.~\ref{fig:potentials} to calculate the KER associated with the KECP.  

The KER of the KECP is calculated as follows.  
The direct non-adiabatic deprotonation KER (if the wavepacket stays on the blue pathway in Fig.~\ref{fig:potentials}) is 6 eV; that is the difference between the final potential energy and the initial potential energy at the Franck-Condon region (C-H stretch $\approx$ 1 \AA.)   
(For now we will ignore any changes to the KER due to rotational kinetic energy, since this will primarily result in a spreading of the KER.)
The KECP, shown by the green dotted line in Fig.~\ref{fig:potentials}, experiences both stimulated emission and a higher asymptotic energy limit compared to the non-adiabatic pathway.  
This results in a total 2 eV decrease in KER compared to the direct dissociation (1 eV from the energy of the photon and 1 eV from the difference in energy between the two final states.)
This corresponds to approximately 4 eV for 1300 nm light, which is the energy of the shoulder in Fig.~\ref{fig:DP_both}.

In comparison, the KER for dashed yellow pathway in the upper panel of Fig.~\ref{fig:potentials} is 6 eV, since the asymptotic limit between the two states is approximately the same as the photon energy of the 1300 nm field. 
Finally, a 5 eV KER release is possible if portions of the wavepacket either change follow the adiabatic pathway, or if portions of the wavepacket experience stimulated emission before the CI, but then stay on the same adiabatic state after the CI.  
This is in contrast to the KECP that requires that the wavepacket experiences non-adiabatic coupling after the CI crossing.  
This analysis leads to the conclusion that there is only one pathway that results in a KER of 4 eV, and that is the KECP.  

The approximate fraction of the population that follows the KECP in Fig.~\ref{fig:DP_both} is 3.6 $\pm$ 1.4 \%.
Such a low probability is quite striking.  
It indicates that these two processes, dipole coupling and non-adiabatic motion, are in a sense, incompatible.  
In other words, if the wavepacket is likely to non-adiabatically transfer through a CI, it is unlikely to experience significant dipole coupling near the CI, and vice versa.    
This small size of this effect is emphasized by comparing to other signals in the spectra such as the large increase in ionization and coupling from higher lying states which shifts the KER spectra towards higher energies.  

Some simple estimates can illuminate the underlying physics that gives rise to these control results. 
The control field experiments are performed in a single photon coupling regime with a field strength of 0.01 au, and a dipole matrix element of approximately 0.2 au.
The Rabi oscillation period is about 76 fs.
The wavepacket will completely cross the CI in 15 fs, therefore the wavepacket spends relatively little time at the point of resonance, reducing the amount of population that follows this pathway.  

This observation highlights how these two coupling terms interact.  
Dipole coupling, an adiabatic process, and non-adiabatic coupling, a non-adiabatic process, may not be simultaneously strong, and a control strategy that requires both couplings is relatively suppressed.
One could increase the amount of time the wavepacket is near the CI by replacing the protons with deuterons to increase the mass, but this also reduces the probability of non-adiabatically crossing the CI, which depends on atomic motion.  
One could also increase the field strength of the control field, but this risks moving from linear regime into a non-linear regime where other effects may become more important.  
The size of this effect is therefore ultimately determined by the dipole moment strength.  

\section{TDSE}
\label{sec:TDSE}

To more conclusively demonstrate that the small effect size we observe is due to the two competing coupling terms, we run a Time Dependent Schrodinger Equation (TDSE) solver to simulate our control experiment.  
This allows us to see if other effects could be causing a reduction in the effect size, such as the dynamic Stark shift, or if the 1300 nm control field was incapable of producing a strong KECP compared to other energies.  

The TDSE simulation we used is similar to the one presented in Ref.~\cite{liekhus-schmaltz_coherent_2016}.  
In brief, the split-operator method is used to propagate the wavepacket in the diabatic basis.
The wavepacket is then rotated to the adiabatic basis where the dipole operator is applied since we calculate the dipole operator in the adibatic frame.
We also include an absorbing boundary defined by 
\begin{equation}
    V_{abs}(x)=i\hbar \ln(\sin^{0.05}(\pi*x/10)),
\end{equation}  
where $x$ is the length of the C-H bond in Angstroms. 
This means that the simulation is effectively 7 \r{A} large.  
In our experiment, the laser is not phase stabilized, so results are averaged over 10 different carrier envelope phases for both the control field and the ionizing field.
We also include the dynamic Stark shift induced by the intense ionizing field in the adiabatic frame. 

To simplify our simulation, we only consider the $^3\Pi_u$ state, which has a conical intersection with the $^3\Sigma_g^+$ state at a carbon-proton distance of about 5 atomic units \cite{gaire_photo-double-ionization_2014}.  
This is the lowest state that deprotonates, and double ionization via multiphoton ionization will most likely lead to this deprotonation process.  
In addition, the $^3\Pi_u$ state has a single crossing with another triplet state, so there is only one kinetic energy pathway.

The potential energy curves, dipole terms, and polarizability were calculated along the C-H stretch coordinate of one proton using a Fractional-Occupation Molecular Orbital Complete-Active-Space Configuration Interaction (FOMO-CASCI) level of theory with an active space that included all valence (8) electrons and 8 active orbitals \cite{slavicek_ab_2010}.
The 6-31g* basis set was used and the electronic temperature parameter in fractional-occupational scheme was chosen to be 0.5 atomic units.  
The FOMO-CASCI parameters were chosen to reproduce MR-CISD results reported in Ref.~\cite{gaire_photo-double-ionization_2014}, following the same procedure as reported in Ref.~\cite{yang_imaging_2018}.  
All electronic structure calculations were performed with the TeraChem software package \cite{moreira_toward_2009}.

In order to preserve the essential topological features of the CI, we ran a 2D simulation with the computed C-H potential for one coordinate and a harmonic potential for the other coordinate.  
This is also the direction for linear coupling in the diabatic electronic basis.
We assume that the dipole matrix is independent of the this coordinate.
The simulation uses the mass of a proton to represent the proton moving during deprotonation. 
The potential was chosen so that the wavepacket is non-spreading.
The width of the wavepacket (0.11 \r{A}) corresponds to the $\sim$1 eV width in the field free KER.  
Since the laser pulses are much longer than the deprotonation time, we assume the field intensity is constant during the simulation, and we use the peak intensity of both the 266 nm and IR pulses, which is approximately 0.01 au (See Section \ref{sec:Experiment}.)  
We use crossed and linearly polarized ionizing and control fields in our simulation because this is polarization condition that produces the strongest control.

To simplify the interpretation of our simulation, we consider the case when the non-adiabaticity is large so that the field free decay always follows the non-adiabatic pathway, and so we set the linear coupling to zero.  
In addition, we turn off the dipole coupling after the conical intersection.
Both of these conditions ensure that the wavepacket can only follow the blue (solid) or green (dotted) pathways in Fig.~\ref{fig:potentials}. 
The KECP probability is therefore given by the excited state population, since no population can arrive at the excited state without taking the KECP.  
Since the deprotonation in acetylene can proceed either adiabatically or non-adiabatically, we expect that this simulation will overestimate the population that takes the KECP.
Nevertheless, we can use this simulation to examine how our choice of photon energy and the dynamic Stark shift affect the KECP.

\begin{figure}
\includegraphics[width=\columnwidth]{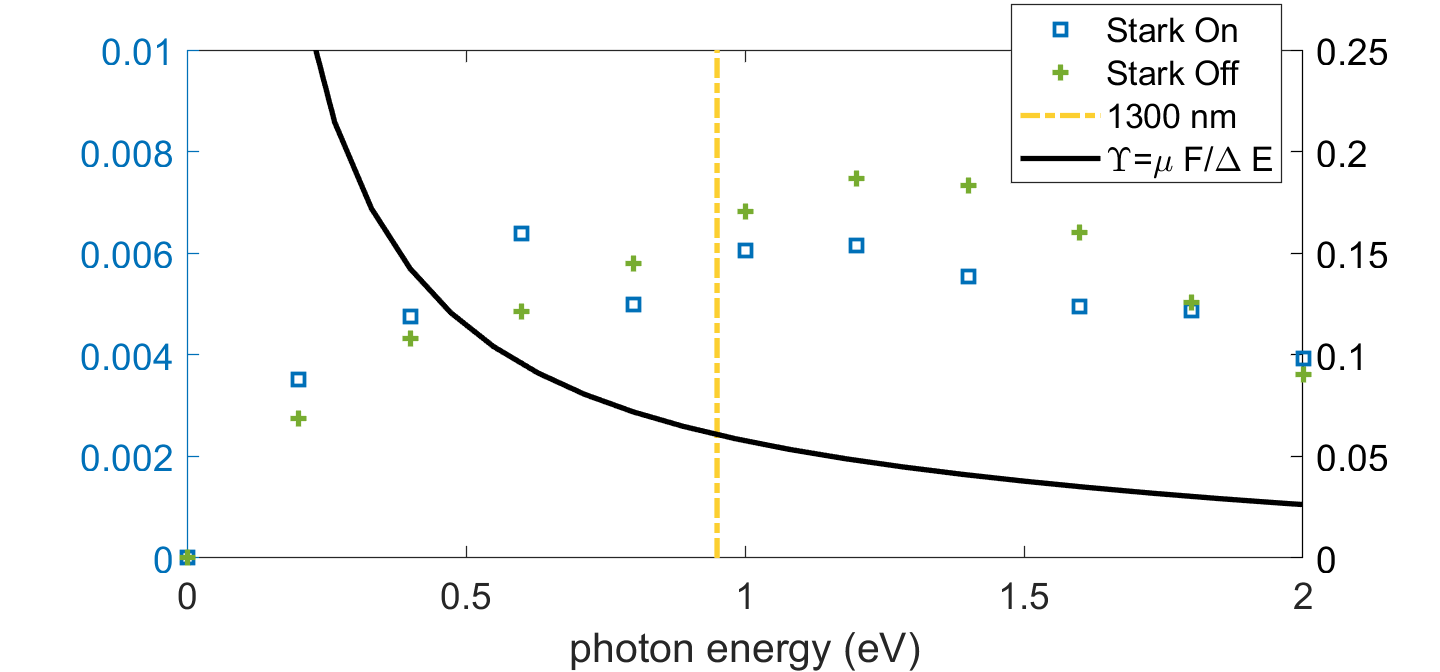}
\caption{TDSE simulation results.  In blue squares and green plus signs are the excited state population for several photon energies at the end of our TDSE simulation.  
Their values are shown with the left hand y-axis.
The simulation is constructed so that the population in the excited state is the population that takes the KECP.  
Immediately apparent is that the relative population that follows the KECP is extremely low, less than a single percentage for all photon energies.  
The blue squares show the population in the excited state with the dynamic Stark shift on, while the green plus signs show the data with the dynamic Stark off.  
By comparing the two at our control energy, we see that the dynamic stark shift does not play a large role in our experiment since the KECP probability is similar for both cases.  
In black is the calculated value for $\Upsilon$, which when it grows, indicates that the excited state populations should decrease.  
Its values are given by the right hand y-axis.
}
\label{fig:compare_params}
\end{figure}

Figure \ref{fig:compare_params} shows how the probability for the KECP changes for different control photon energies and for the dynamic Stark shift.  
The plots show a distinct pattern where the KECP probability is maximized around 1.25 eV.  
A detailed discussion of why this pattern occurs is discussed in detail in Ref.~\cite{liekhus-schmaltz_coherent_2016}.  
In brief, this pattern can be predicted using the Landau-Zener formulation for a non-adiabatic population transfer through a CI and via an adiabatic frequency sweep \cite{wittig_landau-zener_2005,zener_clarence_non-adiabatic_1932,landau_theory_1932}.
This formulation predicts that the KECP probability should increase until the region of dipole coupling and non-adiabatic coupling significantly overlap one another.  
In Ref.~\cite{liekhus-schmaltz_coherent_2016}, we proposed that the ratio of the Rabi frequencies at the CI and at resonance provides a measure for evaluating the proximity of the CI to the point of resonance.  
For reference, we include this measure here.  
Specifically, that ratio for a particular laser frequency, $\omega_L$, transition dipole, $\mu$, and field strength $F$ is:
\begin{equation}
\begin{split}
    \frac{\bar{\Omega}|_{CI}}{\Omega} & = \frac{\sqrt{(\frac{\mu F}{\hbar})^2+(\omega_L)^2}}{\frac{\mu F}{\hbar}} \\
    & = \sqrt{1+\Big(\frac{\Delta E}{\mu F}\Big)^2},
\end{split}
\end{equation}
where $\Delta E = \hbar \omega_L$ is the detuning at resonance.  
In order to ensure that $\nicefrac{\bar{\Omega}|_{CI}}{\Omega}$ is large, meaning that most of the resonant population transfer is complete at the CI, the quantity 
\begin{equation}
    \Upsilon = \frac{\mu F}{\Delta E}
\end{equation}
should be small.  

In Fig.~\ref{fig:compare_params} we can confirm that the photon energy we chose corresponds to the region where the KECP probability is maximized.  
It is also on the lower energy of this region, meaning the region of resonance transfer is starting to overlap with the CI.  
As well, by comparing the KECP probability for the dynamic Stark shift on and off, we see that the dynamic Stark shift has little effect on the KECP probability for the 1300 nm wavelength.  
This shows that the our experimentally observed, small KECP probability is due in large part to the competing timescales associated with dipole and non-adiabatic coupling, rather than the dynamic stark shift or an ill-chosen control wavelength.

\section{Conclusion}

We have demonstrated the existence of a pathway that allows us to directly measure the interaction between dipole coupling, and non-adiabatic coupling.  
We have found that the interaction between these two couplings is necessarily dictated by the timescales associated with ultrafast, non-adiabatic motion.  
Under conditions when the coupling field is weak enough that it will not ionize, we find that this pathway has a small probability that is strongly dictated by the time the molecule takes to pass through a region of resonance.  

These results have important implications for non-adiabatic control schemes that make use of dipole coupling in the region of a CI.  
The relative timescales associated with the field strength and the speed of the wavepacket will determine the strength of the control.  
As a result, control schemes where dipole coupling and non-adiabatic coupling are in direct competition (where they must both be strong) may be fundamentally limited in their effectiveness for systems without a larger dipole moment.  
Therefore, this class of strictly non-adiabatic control schemes should be conducted with this limit in mind.  
Schemes that use either strong non-adiabatic coupling or strong dipole coupling, but not both will be more successful for conducting control in general.

The experiments described in this paper and their analysis and interpretation were carried out by CLS, assisted by GAM and PHB, and supported by the National Science Foundation
under Grant No.~PHY-1504584.  
Theory and calculations were undertaken by XZ, CLS, and TM with support from DOE (AMOS).  
JPC provided critical advice on all aspects, and is supported by the Atomic, Molecular, and Optical Science program within the Chemical, Geo and Biosciences Division of Basic Energy Sciences of the Department of Energy.

\bibliographystyle{unsrt}
\bibliography{references}
\end{document}